\documentclass[fleqn,usenatbib]{mnras}

\usepackage[T1]{fontenc}

\DeclareRobustCommand{\VAN}[3]{#2}
\let\VANthebibliography\thebibliography
\def\thebibliography{\DeclareRobustCommand{\VAN}[3]{##3}\VANthebibliography}

\usepackage{graphicx}	
\usepackage{amsmath}	
\usepackage{multicol}   
\usepackage{bm}	    	
\usepackage{pdflscape}	

\usepackage{newtxtext,newtxmath}

\graphicspath{{./}{figures/}}

\title[Measurement of AGN dust extinction with \it{}WISE\rm{} data]{Measurement of AGN dust extinction based on the near-infrared flux variability of \textit{WISE} data}

\author[S. Mizukoshi et al.]{
Shoichiro Mizukoshi,$^{1}$\thanks{E-mail: s.mizukoshi@ioa.s.u-tokyo.ac.jp}
Takeo Minezaki,$^{1}$
Shoichi Tsunetsugu,$^{1}$
Atsuhiro Yoshida,$^{1}$
\newauthor
Hiroaki Sameshima,$^{1}$
Mitsuru Kokubo,$^{2}$
and Hirofumi Noda$^{3}$
\\
$^{1}$Institute of Astronomy, Graduate School of Science, The University of Tokyo, 2-21-1 Osawa, Mitaka, Tokyo 181-0015, Japan\\
$^{2}$Department of Astrophysical Sciences, Princeton University, Princeton, New Jersey 08544,USA\\
$^{3}$Department of Earth and Space Science, Graduate School of Science, Osaka University, 1-1 Machikaneyama, Toyonaka, Osaka 560-0043, Japan
}

\date{Accepted XXX. Received YYY; in original form ZZZ}

\pubyear{2021}

\begin{document}
\label{firstpage}
\pagerange{\pageref{firstpage}--\pageref{lastpage}}
\maketitle

\begin{abstract}
We present the measurement of the line-of-sight extinction of the dusty torus for a large number of obscured active galactic nuclei (AGNs) based on the reddening of the colour of the variable flux component in near-infrared (NIR) wavelengths.
We collected long-term monitoring data by \it{}Wide-field Infrared Survey Explorer (WISE)\rm{} for 513 local AGNs catalogued by the \it{Swift/}\rm{}BAT AGN Spectroscopic Survey (BASS) and found that the multi-epoch NIR flux data in two different bands (WISE $W1$ and $W2$) are tightly correlated for more than 90\% of the targets.
The flux variation gradient (FVG) in the $W1$ and $W2$ bands was derived by applying linear regression analysis, and we reported that those for unobscured AGNs fall in a relatively narrow range, whereas those for obscured AGNs are distributed in a redder and broader range.
The AGN's line-of-sight dust extinction ($A_V$) is calculated using the amount of the reddening in the FVG and is compared with the neutral hydrogen column density ($N_{\rm{}H}$) of the BASS catalogue.
We found that the $N_{\rm{}H}/A_V$ ratios of obscured AGNs are greater than those of the Galactic diffuse interstellar medium (ISM) and are distributed with a large scatter by at most two orders of magnitude.
Furthermore, we found that the lower envelope of the $N_{\rm{}H}/A_V$ of obscured AGNs is comparable to the Galactic diffuse ISM. 
These properties of the $N_{\rm{}H}/A_V$ can be explained by increase in the $N_{\rm{}H}$ attributed to the dust-free gas clouds covering the line of sight in the broad-line region.
\end{abstract}




\section{Introduction}
\label{sec:intro}

The unified model \citep[e.g.][]{Antonucci93} has been interpreted as the structure of an active galactic nucleus (AGN) in which the dusty torus surrounds the supermassive black hole and accretion disc.
The dusty torus is important for explaining the differences in optical spectra of AGNs, i.e., the Seyfert type, caused by a difference at the viewing angle to the AGN with the common structure.

The dusty torus can function as a mass reservoir for the central engine, supplying dust and gas.
However, the radiation from the central engine will blow out most of the accreting materials into the circumnuclear region or intergalactic space as the AGN feedback \citep[][for review]{Fabian12}. 
This radiation is considered to blow out not only the accreting material but also the interstellar medium in its host galaxy and finally suppress its star formation activity \citep{Combes17,Harrison17}, in addition to the growth of the central black hole.
The feeding and feedback processes between the AGN and its host galaxy may account for their co-evolution over cosmic time, thus resulting in correlations between AGN and host--galaxy properties \citep[e.g.,][]{Kormendy95,Magorrian98,Kormendy13}.


\cite{Fabian09} suggested that radiation pressure on dusty gas in an AGN depends on the dust abundance, hence it is one of the important parameters to yield some hints about the co-evolution.
The optical extinction $A_V$ characterises the amount of dust in the line of sight, which is commonly attributed to the dusty torus.
Optical spectroscopic analyses based on the flux ratios of two or more optical emission lines \citep[e.g., Balmer decrement,][]{Baker38,Ward87,Gaskell17} or the luminosity ratio of the broad H$\alpha$ line and the hard X-ray \citep{Shimizu18} have been performed to measure the $A_V$ of the dusty torus for more than ten thousands of less-obscured AGNs \citep[e.g.,][and citation therein]{Jun21}.
However, because of strong obscuration in the optical band, these methods based on optical observations are difficult to apply to obscured AGNs.

Emission lines in the near-infrared (NIR) band (e.g., the Paschen series) have been used to estimate the $A_V$ of several dozens of obscured AGN targets \citep[e.g.][]{Ward87,Maiolino01,Schnorr-Muller16}.
Alternatively, \cite{Burtscher16} measured the colour temperature of NIR continuum emission for 29 AGNs---about a half of which are obscured AGNs---by the spectral fitting for the $K$-band spectrum.
They considered the average temperature measured for type-1--1.9 AGNs as the "intrinsic" value, without any intervening dust extinction, and estimated the $A_V$ of AGN samples by measuring the temperature decrease from the intrinsic value.
They examined the ratio of the neutral hydrogen column density $N_{\rm{H}}$ that was derived from X-ray spectral analysis (see Section \ref{subsec:BASS sample}) and the $A_V$, and reported that $N_{\rm{H}}/A_V$ is usually larger than the typical value of the Galactic diffuse interstellar medium (ISM) with large scatters, by at most two orders of magnitude, as has been presented in other studies \citep[e.g.][]{Maiolino01,Maiolino01b,Imanishi01}.
\cite{Xu20} derived the $A_V$ of 175 AGNs using the strength of the 9.7 $\mu$m silicate feature and presented that the silicate strength is weakly correlated with $N_{\rm{}H}$ with large scatters.

\cite{Winkler92} presented the optical flux time variation of an AGN as another clue for measuring the $A_V$ without spectroscopic analyses.
They calculated the flux variation gradient (FVG) using the monitoring data from $UBV$ bands, which is the ratio of the amplitude of flux variations in two different bands. 
They used regression analysis on the flux--flux plot to measure the FVG, which takes the flux data from two bands at the same epoch on the vertical and horizontal axes. 
They found that the FVGs for unobscured AGNs are nearly \textcolor{black}{the same value in every target}, which can be considered as the intrinsic FVG.
The $A_V$ for each target was then calculated by measuring the reddening of the FVG relative to
the intrinsic FVG.
This method has the advantage of ignoring host-galaxy emission because it only uses the variable components of the optical flux that are attributed to the AGN.

The flux of the NIR continuum emission is also time variable, responding to that of the UV-optical continuum emission from the accretion disc \citep[e.g.,][]{Suganuma06,Koshida14,Minezaki19,Lyu19,Yang20}.
Such NIR emission is considered to be emitted particularly from the innermost region of the dusty torus, in which hot dust is heated to 1000--2000 K by the strong emission from the central engine \citep[e.g.,][]{Barvainis87,Lyu17,Baskin18}.
\cite{Glass04} investigated the colours of the variable NIR flux components of 41 AGNs---six are obscured AGNs---using monitoring data in \it{}JHKL\rm{} bands.
He demonstrated that the intrinsic NIR FVGs fall within moderately narrow ranges as for the optical FVG reported by \cite{Winkler92}.
Accordingly, \cite{Glass04} mentioned that the FVG in NIR bands may be useful to estimate the dust extinction due to the outer torus for obscured AGNs, and applied his method to NGC1068.

In this study, we used NIR monitoring data from two bands to obtain the NIR FVGs of nearby obscured and unobscured AGNs.
We estimated their line-of-sight optical extinction $A_V$ from the reddening of the NIR FVGs.
Furthermore, by comparing these $A_V$ to $N_{\rm{H}}$ determined using X-ray spectral analyses \citep{Koss17}, we examined the correlation between them and compared it with the typical $N_{\rm{}H}/A_V$ of the Galactic diffuse ISM.
This paper is organised as follows:
We describe the data, the AGN catalogue, and our sample selection in Section \ref{sec:data analysis}.
Section \ref{sec:data correlation and the variable color} presents the flux--flux plot and the derivation of the NIR FVG in detail. 
We also present the results of our samples' NIR FVG properties here.
Section \ref{sec:dust extinction} explains how to convert the  NIR FVG to the $A_V$ and demonstrates the results for our samples.
We compare our results with the typical $N_{\rm{}H}/A_V$ of the Galactic diffuse ISM in Sections \ref{sec:data correlation and the variable color} and \ref{sec:dust extinction}.
In Section \ref{sec:discussion}, after comparing our $A_V$ estimate with other studies, the possible scenarios for explaining the distribution of the $N_{\rm{}H}/A_V$ are discussed.
We then provided our summary in Section \ref{sec:conclusion}.
Throughout this study, we adopted the cosmology $H_0=70\ \mathrm{km\ s^{-1}\ Mpc^{-1}}$, $ \Omega_0=0.30$, and $\Omega_{\Lambda}=0.70$.

\section{targets and data}
\label{sec:data analysis}


\subsection{BASS AGN catalogue}
\label{subsec:BASS sample}

Our targets are obtained from the catalogue of the BAT AGN Spectroscopic Survey \citep[BASS,][]{Koss17,Ricci17b}, which was originally provided in the \it{Swift}\rm{}/BAT 70-month catalogue \citep{Baumgartner13}.
This catalogue contains 836 nearby AGNs detected using the \it{Swift}\rm{}/BAT 14–195 keV band with a redshift peak of $z\sim0.05$.

\cite{Koss17} analysed the optical spectra of about 77\% of the BASS AGNs and classified almost all of them as Seyfert galaxies.
They classified 539 of them into Seyfert types, whereas the BASS website\footnote{\url{https://www.bass-survey.com}} affords the Seyfert types of 594 AGNs.
The latter's Seyfert types are used in this study.

In the BASS AGN catalogue, the values of $N_{\rm{}H}$ were primarily  obtained by X-ray spectral analysis using not only 14–195 keV data but also soft X-ray data at 0.3–10 keV, which was obtained from \it{}XMM-Newton\rm{} \citep{Jansen01}, \it{}Chandra\rm{} \citep{Weisskopf00}, \it{}Suzaku\rm{} \citep{Mitsuda07}, or \it{Swift}\rm{}/XRT \citep{Burrows05}.
In the $N_{\rm{}H}$ measurement, the Galactic absorption was considered in advance using the value from the H{\sc i} maps of \cite{Kalberla05}.
The $N_{\rm{}H}$ of AGNs were then measured in which photoelectric absorption and Compton scattering were considered using the ZPHABS and CABS models, respectively \citep{Ricci17b}.
The lower limit measured by the analysis was a $\log N_{\rm{H}}\ [\rm{cm}^{-2}] = 20$  \citep{Koss17,Ricci17b}, which is lower than the typical value of the Galactic diffuse ISM.
\cite{Ricci17b} estimated the $N_{\rm{}H}$ of 75 Compton-thick AGNs using torus modelling to precisely determine them.
We note that \cite{Burtscher16} used the $N_{\rm{}H}$ in the BASS AGN catalogue for their investigation as in this study.


\subsection{Details about WISE data}
\label{subsec:WISE details}

We used the NIR photometric data obtained by the \it{}Wide-field Infrared Survey Explorer\rm{} \citep[\it{WISE}\rm{},][]{Wright10}. 
\it{WISE}\rm{} was launched in 2009 and performed its cryogenic all-sky survey for about a year in four bands: $W1$ ($3.4\ \mu$m), $W2$ ($4.6\ \mu$m), $W3$ ($12\ \mu$m), and $W4$ ($22\ \mu$m). 
This all-sky survey is called the ALLWISE programme.
\it{WISE}\rm{} was reactivated without cryogen after a two-year hibernation period to conduct an all-sky monitoring survey in the $W1$ and $W2$ bands to examine near-Earth objects \citep[NEOWISE,][]{Mainzer11,Mainzer14}.
Currently, \it{WISE}\rm{} provides the all-sky monitoring data in $W1$ and $W2$ bands spanning for more than ten years, which comprises at most three epoch data from ALLWISE and typically 15 epoch data from NEOWISE.
Each epoch data comprises typically a few dozen of single-exposure data.

\begin{figure*}
    \includegraphics[width=\linewidth]{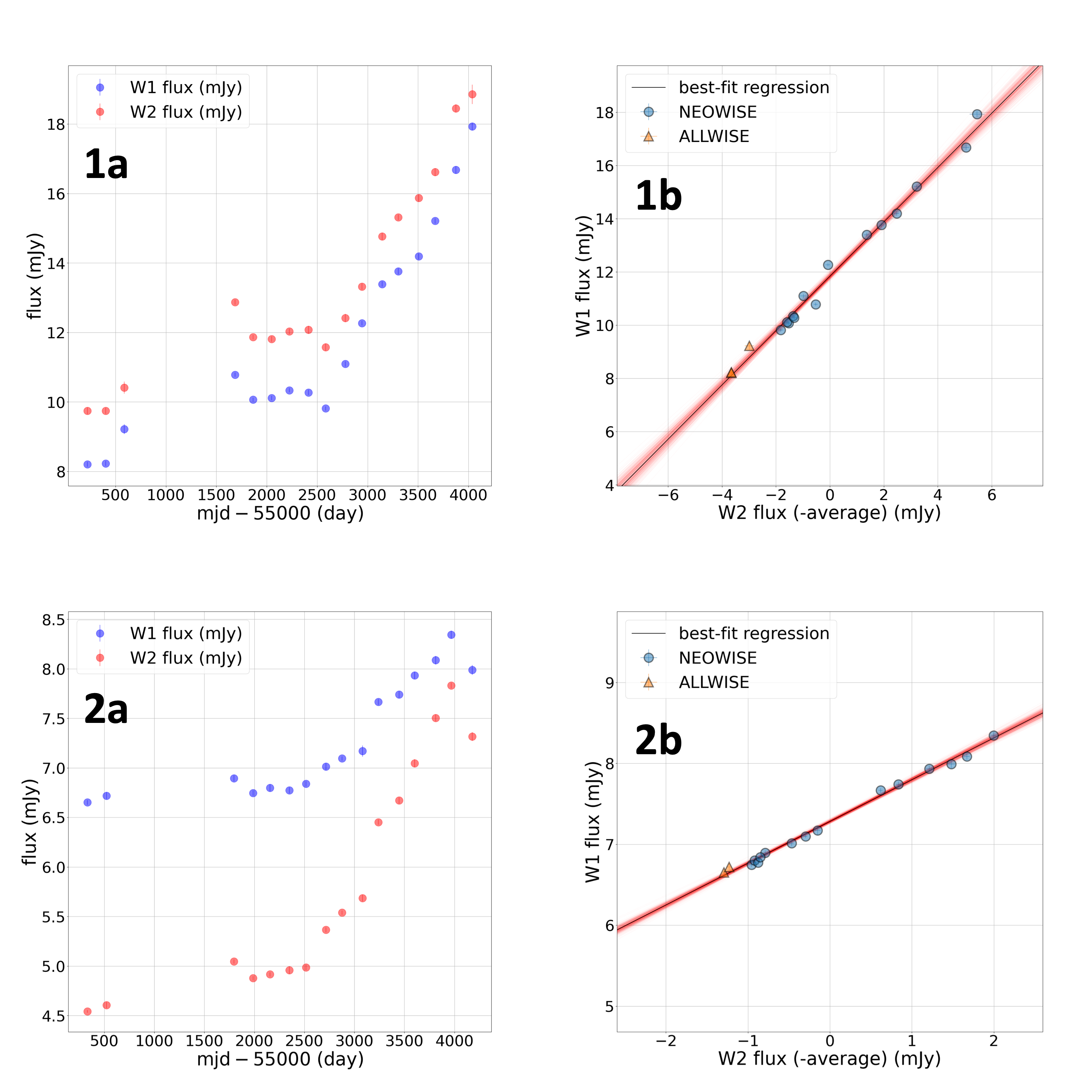}\par 
\caption{\it{}1a\rm{}: The NIR light curve of LEDA 126226, as an example of the typical type-1 AGN.
The blue and the red circles indicate the $W1$- and $W2$-band fluxes, respectively.
The error bar of each data point represents $\pm1\sigma$ error of the flux data. 
\it{}1b\rm{}: The $W1$-band flux to $W2$-band flux plot of LEDA 126226 as an example.
The blue circles indicate NEOWISE data and the orange triangles indicate ALLWISE data.
The black solid line indicates the best-fit regression line for the data points,
and the red thin lines show the 400 samples from the posterior distribution of the regression.
\it{}2a\rm{}: The same figure as 1a for LEDA 2730634 as an example of the typical type-2 AGN.
\it{}2b\rm{}: The same figure as 1b for LEDA 2730634 as an example of the typical type-2 AGN.}
 \label{fig:example}
\end{figure*}

Typically, \it{WISE}\rm{} observes each target once every six months.
We averaged the photometric data in each epoch after excluding outliers that deviated from the average by more than $3\sigma$ of the distribution of the other flux data in the same epoch.
We excluded monitoring data points from the epoch that did not have $W1$- or $W2$-band photometric data.
The analysis of the NIR FVG described in the following sections demonstrates that it falls in the range corresponding to the power-law spectrum of $F_{\nu}\propto\nu^{\alpha}$ with $-4 <\alpha<0$.
We, therefore, adopt the zero magnitude flux densities of 306.682 and 170.663 Jy for the $W1$ and $W2$ bands, respectively, which are those for $F_{\nu}\propto\nu^{-2}$ \citep{Wright10}.
In this study, the uncertainty of the derived flux density because of the difference of zero magnitude flux density is at most about 3\% \citep{Wright10,Jarrett11}, which is smaller than the typical uncertainty of the NIR FVG in this study.
We corrected for the Galactic extinction in the flux calculation.
We adopted the Galactic extinction for each target from NASA/IPAC Extragalactic Database (NED), which is based on \cite{Schlafly11}.
We demonstrate examples of the light curve using the \it{}WISE\rm{} data in Fig. \ref{fig:example} (1a) and (2a) for the typical type-1 and type-2 AGNs, respectively.

\subsection{Target selection}
\label{subsec:sample selection}

First, we selected 594 AGNs in the BASS catalogue with the Seyfert type.
We then excluded 25 AGNs that are classified as blazars in the Roma Blazar Catalogue \citep[BZCAT,][]{Massaro09}.
Targets with the Galactic extinction $A_V(\mathrm{Gal.})>2$ mag were excluded to decrease the uncertainty of the estimated $A_V$ caused by that of the Galactic extinction.
Typically, obscured AGNs have $\log N_{\rm{}H}\ [\mathrm{cm^{-2}}]\gtrsim22$, which corresponds to $A_V\gtrsim5$ mag assuming the typical $N_{\rm{}H}/A_V$ of the Galactic diffuse ISM \citep{Predehl95,Nowak12}.
We excluded 26 targets here.

Next, we set the target redshift to $z < 0.5$ so that the thermal emission from the hot dust in the innermost region of the dusty torus dominated the NIR observed fluxes.
Because most of the BASS AGNs are nearby sources, only two AGNs were excluded.

We then selected targets with good-quality \it{WISE}\rm{} data, which are defined by the data quality flags as good frame quality (\tt{}qual\_frame$\,\geq5$\rm{}), good signal-exposure image (\tt{}qi\_fact$\,=0.5,\,1$\rm{}), no charged particle hits (\tt{}saa\_sep$\,\geq5$\rm{}), no moon effect (\tt{}moon\_masked$\,=0$\rm{}), and no effect from artifacts (\tt{}cc\_flags$\,=\,$’0000’\rm{}).
Here we excluded the targets with no good-quality data from our samples.
Moreover, we excluded targets that had at least one saturated pixel in more than 10\% of the observational data across all epochs, whereas we used \it{WISE}\rm{} profile-fitting photometry, which used only unsaturated pixels.
We excluded 28 targets here in total.
Therefore, we selected 513 AGNs for subsequent analysis.
Although some targets lacked usable data from the ALLWISE programme, we did not exclude them because our NIR FVG analysis could be adequately conducted only with the data from the NEOWISE programme.

Figs. \ref{fig:example} (1b) and (2b) show the $W1$-band flux to $W2$-band flux plot for the targets, LEDA 126226 and LEDA 2730634, as examples of typical type-1 and type-2 AGNs, respectively.
In the horizontal axis of the flux--flux plot, we consider the difference from the mean $W2$ flux value to approximately minimise the uncertainty of the intercept of the vertical axis. 
The data points spread in the flux--flux plot were then fitted using the linear relation,

\begin{equation}
         f_{W1} = \alpha + \beta\,(f_{W2} - \langle f_{W2}\rangle) , 
\end{equation}
where $f_{W1}$ and $f_{W2}$ are $W1$- and $W2$-band fluxes, respectively, and $\langle f_{W2}\rangle$ is the average of the $W2$-band fluxes.
We set the intercept $\alpha$ and slope $\beta$ as free parameters, and the NIR FVG for the target is estimated as $\beta$ of the best-fit linear regression.

Before performing linear regression analysis, we calculated the correlation coefficient $r$ between the $W1$ and $W2$ flux data to evaluate the strength of the correlation between flux variations in $W1$ and $W2$ bands.
We reported that the flux variations in the two  NIR bands were highly correlated for the majority of the targets ($r \geq 0.9$ for 83\% of the 513 targets), which is consistent with the results of \cite{Glass04}.
Moreover, our result is based on an order of magnitude larger sample, including multiple obscured AGNs.

The targets having a high correlation coefficient were subjected to linear regression analysis.
However, there are some targets with high correlation coefficients for which the NEOWISE programme data points are clustered in a small area on the flux--flux plot, thus suggesting small flux variation and weak correlation for the NEOWISE data. 
A few data points from the ALLWISE programme located separately may produce a strong correlation in such cases.
In this study, we focus on the flux variation of the thermal radiation on a timescale of a few to ten years; hence, our suitable targets should demonstrate the clear flux variation even in the NEOWISE data alone.
Consequently, we excluded targets with weak correlation for the NEOWISE data based on the correlation coefficient, calculated using only data from the NEOWISE programme ($\equiv r_{\rm{}NEO}$).
466 AGNs were selected for which both $r$ and $r_{\rm{}NEO}$ were larger than 0.7 because 0.64 is the value of the correlation coefficient indicating a 99\% confidence level for a sample size of 15.
Many targets eliminated here show not only the small flux variation but also the large $N_{\rm{}H}$ with the peak of $\log N_{\mathrm{H}}\ [\mathrm{cm^{-2}}]\sim23.5$.
This indicates that these targets may be heavily obscured sources.

\section{near-infrared flux variation gradient}
\label{sec:data correlation and the variable color}

\subsection{Linear regression analysis of the flux--flux plot}
\label{subsec:flux-flux plot}

We performed linear regression analysis on the flux--flux plots of the 466 targets.
Because data points in the flux--flux plot have errors on both axes, we used a Python port of a Bayesian linear regression routine called Linmix\_err in IDL developed by \cite{Kelly07}, called Linmix\footnote{Linmix: \url{http://linmix.readthedocs.org/}}, which normally incorporates distributed errors in dependent and independent variables.
Linmix assumes the distribution of independent variables with multiple Gaussians, and we herein set the number of Gaussian $K$ as $K=2$, which is the smallest number we can set to make Linmix work properly.
In this study, we included the intrinsic scatter in our calculation.

Figs. \ref{fig:example} (1b) and (2b)  show the best-fit regression line and samples from the posterior distribution of the regression for each target.
We estimated the NIR FVGs with the good precision of the FVG error ($\equiv\sigma_{\beta}$) being smaller than 0.2 for the 463 targets.
This represents more than 90\% (463/513) of the local non-blazar BASS AGNs with known Seyfert type, small Galactic extinction, and good \it{}WISE\rm{} data.
We selected these 463 targets to estimate the line-of-sight dust extinction using the NIR FVG described in the following sections.
Table \ref{tab:result} lists the estimated NIR FVGs and errors for the target AGNs.

\begin{figure*}
\begin{multicols}{2}
    \includegraphics[width=\linewidth]{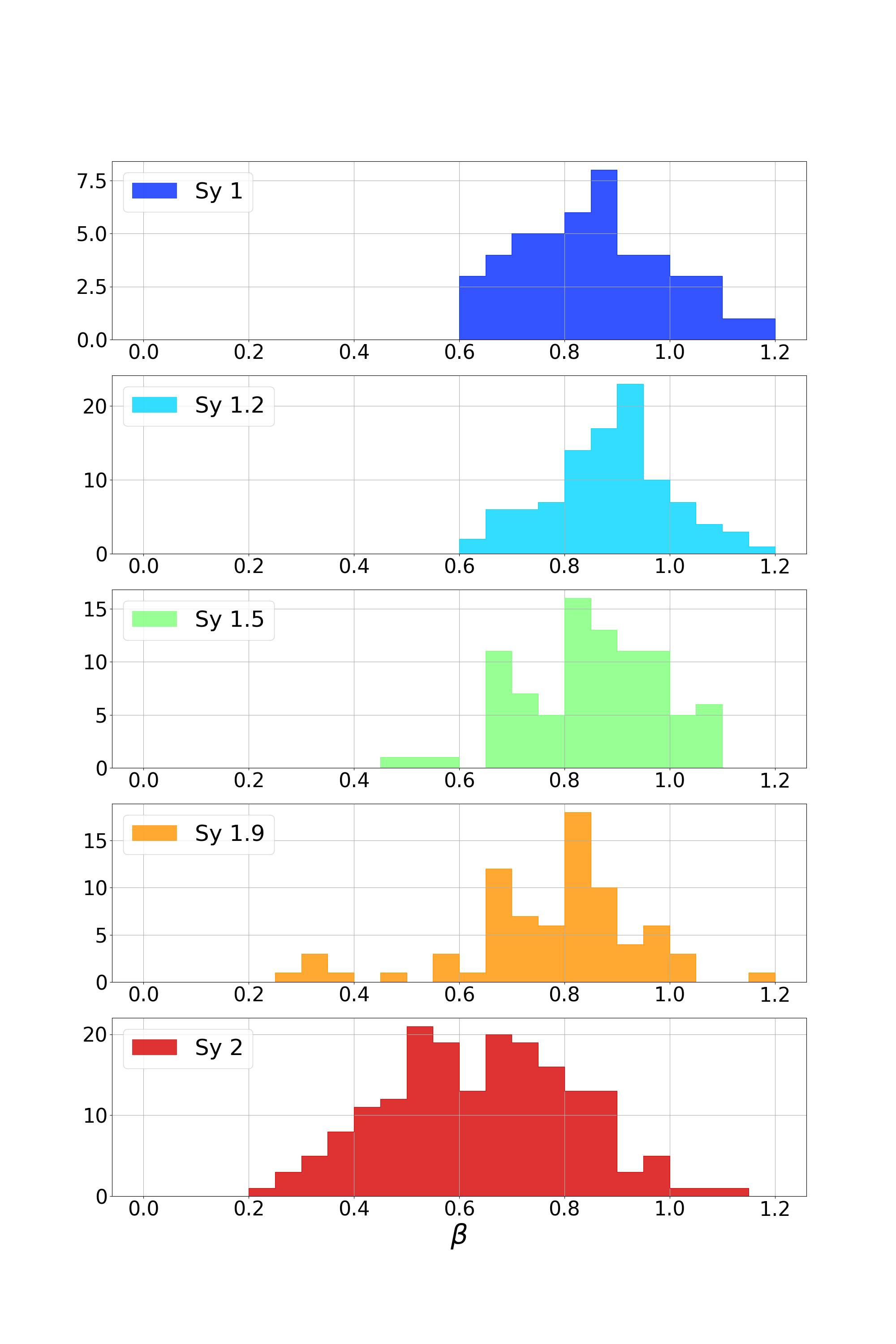}\par 
    \includegraphics[width=\linewidth]{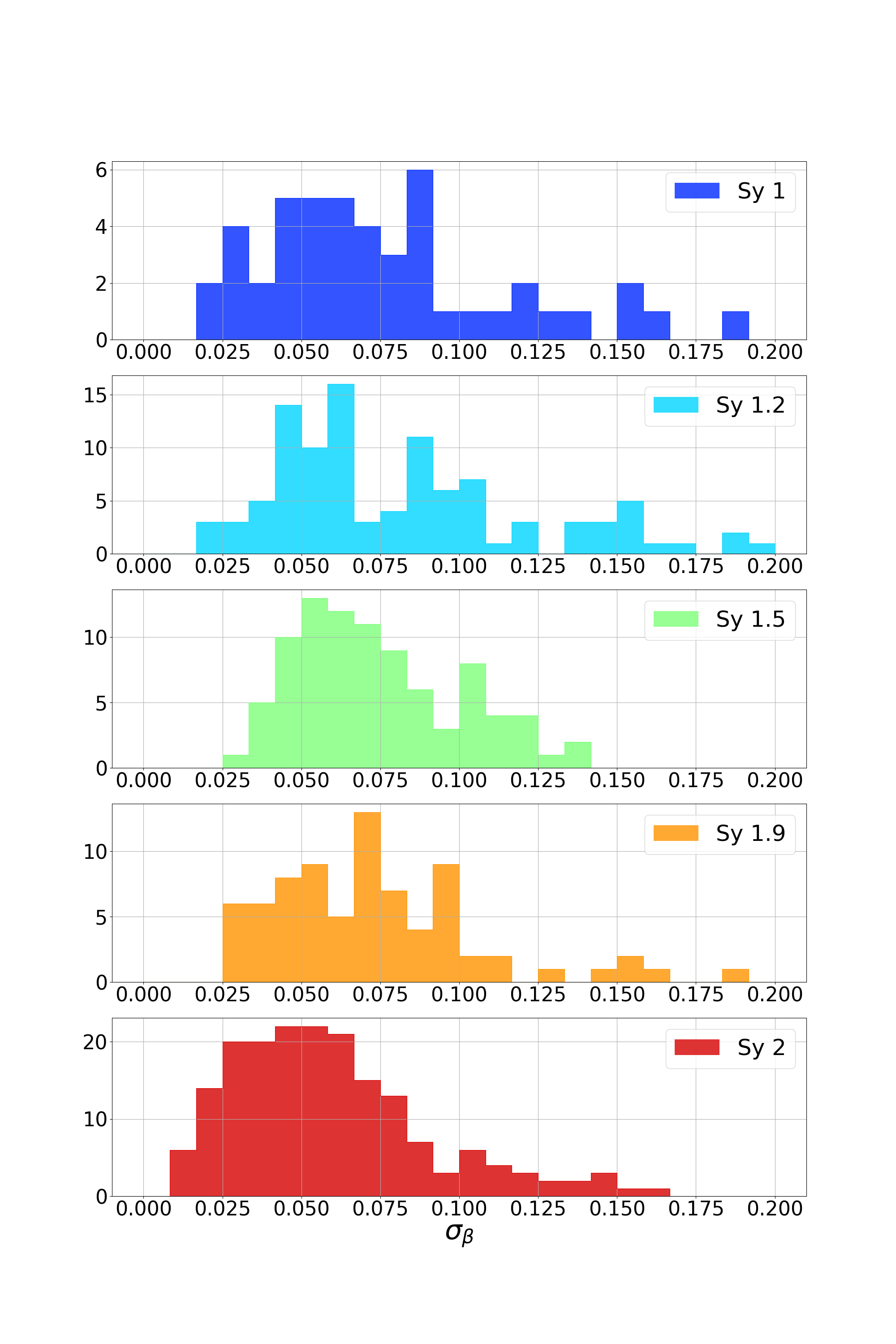}\par 
\end{multicols}
\caption{\it{}Left\rm{}: The histograms of the NIR FVG $\beta$ for the 463 AGN samples.
From top to bottom panels, they represent the histograms
for the Seyfert type-1 (blue), type-1.2 (cyan),  type-1.5 (green),
type-1.9 (orange), and type-2 (red) targets, respectively.
\it{}Right\rm{}: The histograms of the error of the NIR FVG $\sigma_{\beta}$ for the same samples.
The top to bottom panel order and colour code is the same as the left panels.}
\label{fig:hist_final}
\end{figure*}


\subsection{Results on the near-infrared FVG}
\label{subsec:beta characteristics}

Fig. \ref{fig:hist_final} shows the distribution of the NIR FVG $\beta$ (the left panel) for the 463 targets in a sequence by Seyfert type.
These targets comprise 40, 93, 83, 70, and 177 targets of Seyfert type 1, 1.2, 1.5, 1.9, and 2, respectively.
We found that the NIR FVGs for less obscured (type-1--1.5) AGNs were distributed around $\beta\sim0.9$ in a relatively narrow range as reported by \cite{Glass04}.
However, we found a clear reddening trend of the NIR FVGs for type-2 AGNs.
They were broadly distributed in a range of $\beta\sim0.4$--0.9 with many targets at $\beta<0.4$.
Observed flux time variation in the $W1$ band is more likely to be suppressed than that in the $W2$ band because of dust extinction, which makes the regression line flatter or $\beta$ smaller in obscured AGNs. 
The NIR FVG of type-1.9 AGNs is distributed around $\beta\sim0.8$ with a small number of the targets at $\beta<0.5$.

Fig. \ref{fig:hist_final} shows the distribution of the FVG error $\sigma_{\beta}$ (the right panel).
The distribution of the FVG error differs little between the targets of different Seyfert types.
The average FVG error for all targets is $\langle\sigma_{\beta}\rangle = 0.07$ with only 18\% (84/463) of targets having relatively large FVG errors of $\sigma_{\beta} > 0.1$.

\begin{figure*}
    \includegraphics[width=0.7\linewidth]{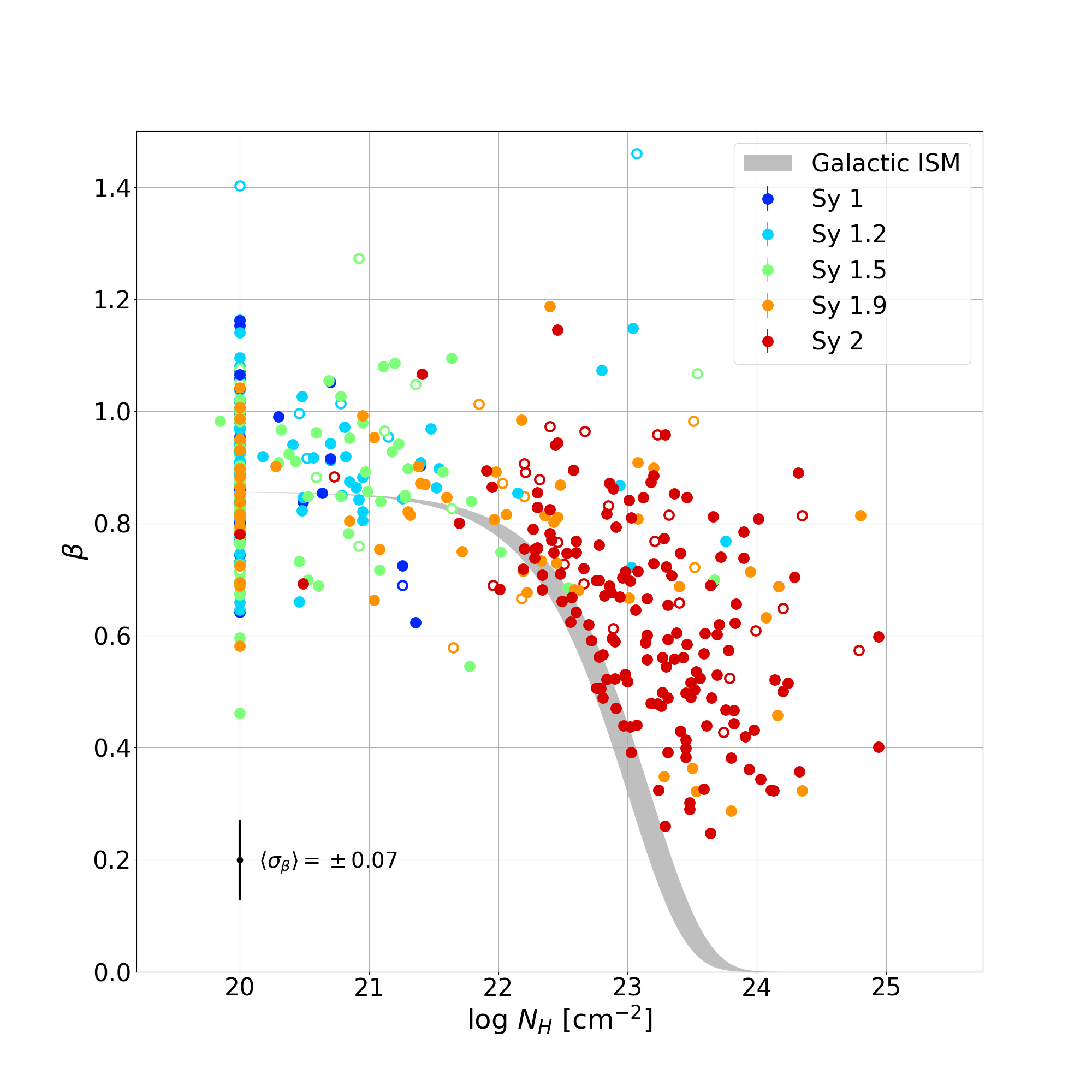}\par 
\caption{The NIR FVG $\beta$ plotted against the line-of-sight hydrogen column density $N_{\rm{}H}$.
The filled circles represent the data points with $\sigma_{\beta}<0.1$, and the open circles represent those with $0.1\leq\sigma_{\beta}<0.2$.
The colours of circles indicate the different Seyfert types of the targets in the same way as in Fig. \ref{fig:hist_final}.
The black segment in the lower left represents the mean $\pm1\sigma$ error of the near-IR FVGs, $\langle\sigma_{\beta}\rangle=0.07$.
The grey band represents the typical relationship between the NIR FVG and the hydrogen column density of the Galactic diffuse ISM (see Section \ref{subsec:Av characteristics}). }
\label{fig:NH_beta}
\end{figure*}

In Fig. \ref{fig:NH_beta}, the NIR FVG is compared with $N_{\rm{}H}$ measured by X-ray absorption.
The colours of the data points are coded in a sequence by Seyfert types.
We successfully obtained the NIR FVGs for both unobscured and obscured AGNs whose hydrogen column densities ranged from $\log N_{\rm{}H}\ [\rm{cm}^{-2}]=20$
(lower limit of the measurement) to $\log N_{\rm{}H}\ [\rm{cm}^{-2}]\sim25$.
We found that the NIR FVGs for less obscured AGNs of $\log N_{\rm{}H}\ [\rm{cm}^{-2}]\lesssim22$ fall in a relatively narrow range. These AGNs are dominated by type-1--1.5 AGNs.
However, we found that the NIR FVGs show clear reddening for obscured AGNs of $\log N_{\rm{}H}\ [\rm{cm}^{-2}]\gtrsim22$, which are dominated by Seyfert type-1.9--2 AGNs.

We note that there is a large difference in the NIR FVG at fixed $N_{\rm{}H}$ for obscured AGNs of $\log N_{\rm{}H}\ [\rm{cm}^{-2}]\gtrsim22$; accordingly, there is a large difference in $N_{\rm{}H}$ by at most two orders of magnitude at fixed NIR FVG for obscured AGNs of Seyfert type 1.9--2.
This indicates that there is a significant scatter in the ratio of dust reddening to the hydrogen column density for obscured AGNs as has been reported \citep[e.g.,][]{Maiolino01,Burtscher16}.
This will be discussed in the following sections.

\section{LINE-OF-SIGHT DUST EXTINCTION}
\label{sec:dust extinction}

\subsection{Calculation of dust extinction from near-infrared FVG}
\label{Av estimation}

The clear trends of reddening of the NIR FVG for obscured AGNs in Figs. \ref{fig:hist_final} and \ref{fig:NH_beta} show that the line-of-sight dust extinction of AGNs can be estimated from the amount of reddening of the NIR FVG.
Here we explain the formula for estimating the dust extinction from  the NIR FVG.

The time-variable components of $W1$- and $W2$-band fluxes, $\tilde{f}_{W1}$ and $\tilde{f}_{W2}$, respectively, are expressed as follows:

\begin{equation}
\tilde{f}_{W1}=\tilde{f}_{W1,0}\times10^{-2/5\times A_{W1}},
\label{eq:2}
\end{equation}

\begin{equation}
\tilde{f}_{W2}=\tilde{f}_{W2,0}\times10^{-2/5\times A_{W2}},
\label{eq:3}
\end{equation}
where $\tilde{f}_{W1,0}$ and $\tilde{f}_{W2,0}$ are the $W1$- and $W2$-band time-variable fluxes  without extinction, respectively, and $A_{W1}$ and $A_{W2}$ are the magnitudes of extinction in the $W1$ and $W2$ bands, respectively.
Taking the logarithm of the ratio of Equations (\ref{eq:2}) and (\ref{eq:3}),
\color {black}
the relationship between the $W1$-band to $W2$-band FVG $\beta=\tilde{f}_{W1}/\tilde{f}_{W2}$, $A_{W1}$, and $A_{W2}$ can be expressed as follows:

\begin{equation}
\log\beta=\log\beta_0-\frac{2}{5}\,(A_{W1}-A_{W2}),
\label{eq:4}
\end{equation}
where $\beta_0=\tilde{f}_{W1,0}/\tilde{f}_{W2,0}$ is the $W1$-band to $W2$-band FVG for the AGNs without extinction.
Equation (\ref{eq:4}) can be expressed using the magnitude of the $V$-band dust extinction $A_{V}$ and the dust extinction curve as follows:

\begin{equation}
\log\left(\frac{\beta}{\beta_0}\right)=-\frac{2}{5}\,(k_{W1}-k_{W2})A_{V},
\end{equation}
where $k_{W1}= A_{W1}/A_{V}$ and $k_{W2}= A_{W2}/A_{V}$.
We assumed the $A_{V}$ as the rest-frame $V$-band extinction in the AGN, while $A_{W1}$ and $A_{W2}$ are the extinction at the rest-frame wavelengths of $\lambda_{W1}/(1+z)$ and $\lambda_{W2}/(1+z)$ at the target redshift of $z$, respectively, where $\lambda_{W1}=3.4\,\mu$m and $\lambda_{W2}=4.6\,\mu$m.
Finally, we can calculate the amount of dust extinction in the AGN based on the reddening of the NIR FVG by using the following equation,

\begin{equation}
A_{V}=-\frac{5}{2(k_{W1}-k_{W2})}\log\left(\frac{\beta}{\beta_0}\right)
\label{eq:5}
\end{equation}

We assumed the standard extinction curve of the Galactic diffuse ISM \citep{Fitzpatrick99} for $k_{W1}$ and $k_{W2}$,
and $R_V=3.1$ to obtain $k_{W1}=0.064$ and $k_{W2}=0.045$, respectively, when $z=0$.
We assumed a foreground screen geometry for the obscurer because the NIR emitting region is much smaller than the parsec-scale outer torus \citep{Burtscher15,Burtscher16,Lyu19,Minezaki19,Noda20,Gravity20,Gamez_Rosas22}.

The uncertainty of the NIR FVG for each target $\sigma_{\beta}$ and that for the AGN without extinction $\sigma_{\beta_0}$ are transferred to that of $A_V$ for the target as follows:

\begin{equation}
\sigma_{A_V}=\frac{5}{2\ln(10)\times(k_{W1}-k_{W2})}\sqrt{\left(\frac{\sigma_{\beta}}{\beta}\right)^2+\left(\frac{\sigma_{\beta_0}}{\beta_0}\right)^2}.
\label{eq:sigma_Av}
\end{equation}


\subsection{Near-infrared FVG for unobscured AGNs}
\label{subsec:variable color estimation}

As discussed in Section \ref{subsec:beta characteristics}, the NIR FVGs for less obscured AGNs fall in a relatively narrow range. 
Then, we estimated the NIR FVG for unobscured AGNs $\beta_0$ to investigate the dust extinction of obscured AGNs.
We selected the NIR FVGs of 95 optically- and X-ray-unobscured AGN targets.
They have Seyfert types of type 1--1.5 and their hydrogen column densities are $\log N_{\rm{}H}\ [\rm{}cm^{-2}]\leq20$.
They have relatively small uncertainties of the NIR FVG ($\sigma_{\beta}<0.1$).

\begin{figure}
 \includegraphics[width=\columnwidth]{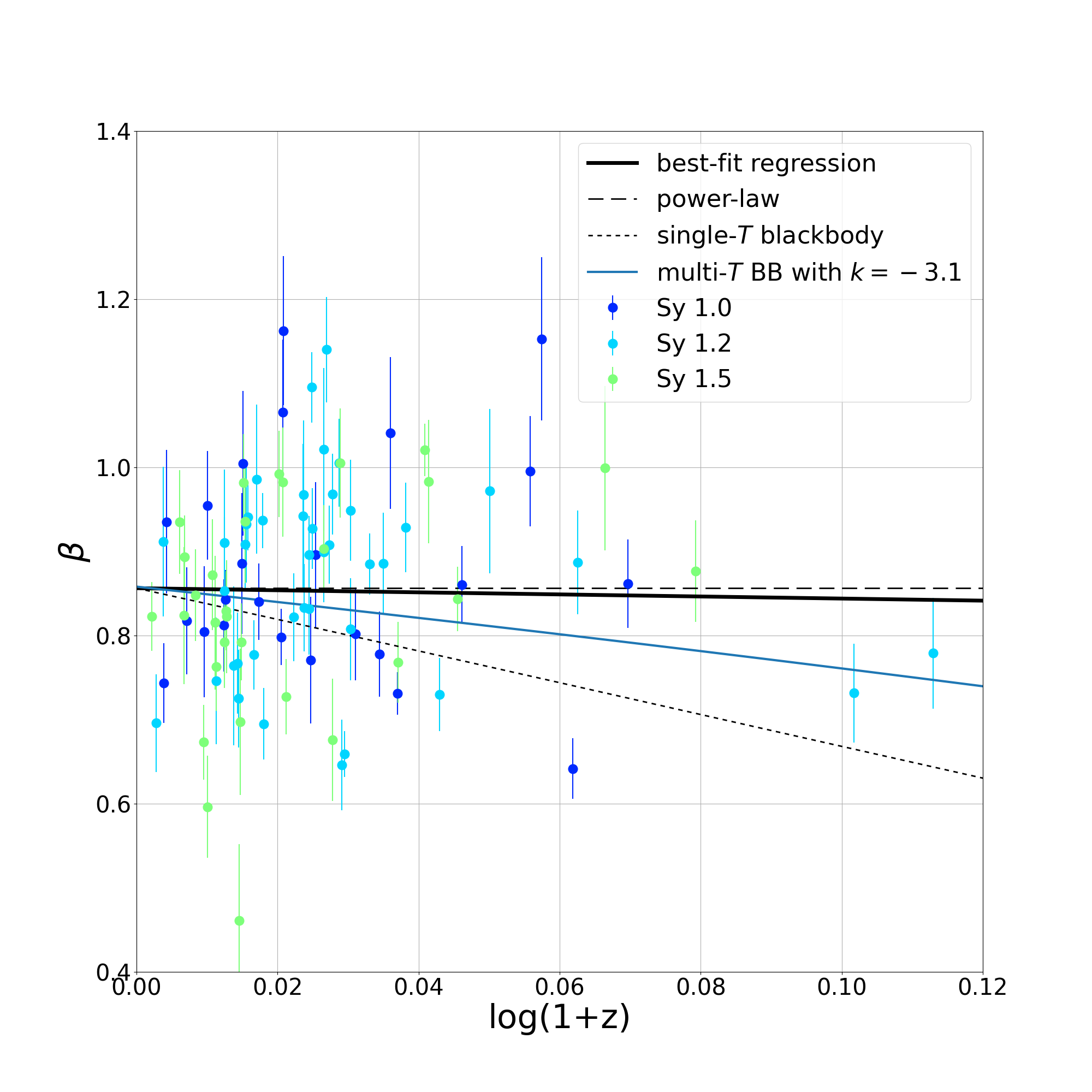}
 \caption{
 The NIR FVG  $\beta$ of the unobscured targets and their redshift. The colours of the data points indicate the different Seyfert types of the targets, in the same way as in Fig. \ref{fig:hist_final}, and the error bars represent $\pm1\sigma$ errors of the $\beta$. The black solid line represents the best-fit regression line to the data. The black dashed line represents the power-law model (power-law index $\alpha=-0.5$), which has the same $\beta$ at $z=0$. The black dotted line represents the single-temperature blackbody model ($T=1077$ K), which has the same $\beta$ at $z=0$. 
 The blue line represents the multi-temperature blackbody model with the maximum temperature of 2000 K and the distribution of the multi-temperature dust of $N(T)\propto T^{-3.1}$.  }
 \label{fig:beta_hist_NH20}
\end{figure}

Fig. \ref{fig:beta_hist_NH20}
shows the NIR FVGs against the redshifts for these unobscured targets.
Their NIR FVGs fall in a relatively narrow range regardless of Seyfert type but appear to be spread more than expected from the measurement errors, indicating some amount of the target-to-target variation in the NIR FVGs for unobscured AGNs, or the intrinsic scatter $\sigma_{\beta_0}$.
To incorporate the possible redshift dependence, we assumed the following relation:

\begin{equation}
\beta_0(z)=\beta_{0}(z=0) + b\,\log (1+z),
\label{eq:beta0}
\end{equation}
and we fitted Equation (\ref{eq:beta0}) to the data by the weighted least squares method, adding the intrinsic scatter of the NIR FVG $\sigma_{\beta_0}$ to the error of the NIR FVG by root-sum-square, $\sqrt{\sigma_{\beta}^2 +\sigma_{\beta_0}^2}$, so that the reduced $\chi^2$ reaches unity.
The best fit parameters are obtained as $\beta_0(z=0)=0.86\pm0.02$ and $b=-0.12\pm0.57$ with $\sigma_{\beta_0}=0.10$.
 Accordingly, we used
\begin{equation}
\beta_0=0.86-0.12\log(1+z),
\label{eq:beta0_result}
\end{equation}

\begin{equation}\sigma_{\beta_0}=0.10,
\label{eq:sigma0}
\end{equation}
for calculating the dust extinction and its error from the NIR FVG in the following sections.

Assuming a blackbody spectrum, $\beta_0(z=0)=0.86$ and $\sigma_{\beta_0}=0.10$ corresponds to the colour temperature of $T=1077\pm140\ \rm{}K$,
which is consistent with that of hot dust considered to be located in the innermost dusty torus \citep[e.g.,][]{Mor09,Netzer15,Honig19,Lyu21}.
$T=1077\pm140\ \rm{}K$ is slightly lower than those estimated for the NIR $H-K$ and $K-L$ FVG colours for the local type-1 AGNs by \cite{Glass04}, which is possibly attributed to the difference in observed wavelengths.
The $W1-W2$ FVG colour tends to be more sensitive to lower-temperature dust thermal emission than the $H-K$ and $K-L$ FVG colours.

As shown in Fig. \ref{fig:beta_hist_NH20}, the NIR FVGs of unobscured AGNs are almost unchanged systematically \color{black} with redshift. 
The FVG is constant regardless of redshift when the spectral energy distribution (SED) of the flux variation is a power-law spectrum, whereas it becomes redder in higher redshift when the SED is a blackbody spectrum of single-temperature dust.
Therefore, such a weak redshift dependence of the NIR FVG suggests that the SED of the flux variation in 3--5 $\mu$m shows a bump much broader than a blackbody spectrum, as is found in the typical SED of quasars \citep[e.g.,][]{Elvis94,Hernan16,Lyu17,Hickox18}. 
This weak redshift dependence of NIR FVG and possible broad bump in the NIR SED of AGNs
\color{black}
can be explained by a composite of thermal radiation from multi-temperature dust. 
By assuming a distribution of the amount of multi-temperature dust as a power-law of temperature $N(T)\propto T^{k}$ with a maximum temperature of 2000 K, the $W1$- to $W2$-band flux ratio of the composite SED of the blackbody radiation from the multi-temperature dust matches the best-fit $\beta_0(z=0)$ of 0.86 when $k\sim-3.1$.
We show the redshift dependence of the flux ratio in Fig. \ref{fig:beta_hist_NH20} with the blue line.
The reddening of the flux ratio \textcolor{black}{of multi-temperature blackbody radiation} is weaker than that of the single-temperature blackbody radiation  in higher redshift, and this result suggests that the SED of the blackbody radiation from multi-temperature dust shows a broader bump in 3--5 $\mu$m than that from single-temperature dust.

\color{black}
Then, we compared the $\beta_0$ with $W1$- to $W2$-band flux ratios derived from composite SEDs of quasars.
The flux ratio from the SED of \cite{Hernan16} is 0.73, which is redder than $\beta_0$ even if the intrinsic scatter $\sigma_{\beta_0}$ is considered.
We also compared the $\beta_0$ with the flux ratio derived from the composite SED of warm-dust deficient (WDD) quasars that are defined by \cite{Lyu17}.
WDD quasars have a broad NIR bump in their SEDs as normal quasars, but they show weaker mid-infrared emission, which is attributed to the reduction of the warm dust component in the dusty torus.
We derived the flux ratio from the composite SED of the WDD quasars as 0.81.
We found that it falls within the intrinsic scatter of $\sigma_{\beta_0}=0.10$ from the best-fit NIR FVG of $\beta_0(z=0)=0.86\pm0.02$, while the $\beta_0(z=0)$ is still bluer than the flux ratio for the WDD quasars.
This result indicates that the variable component of flux in 3--5 $\mu$m are dominated by the thermal emission of hot dust in the innermost region of the dusty torus, which is so compact that the thermal reradiation response to the flux variation of the central engine is larger than that in the outer torus. 
This strengthens the view of a foreground screen geometry for the obscurer to measure the line-of-sight extinction based on the reddening of the NIR FVG, as described in Section \ref{Av estimation}.

\subsection{Results of dust extinction}
\label{subsec:Av characteristics}

Table  \ref{tab:result} lists the dust extinction estimated from the NIR FVG and the error based on Equations (\ref{eq:5}), (\ref{eq:sigma_Av}), and (\ref{eq:beta0_result}).
We estimated the dust extinction of obscured AGNs up to $A_V\sim65$ mag.
The mean uncertainty of $A_V$ of our targets is $\sigma_{A_V}=8.3$ mag. 

\begin{figure*}
    \includegraphics[width=0.7\linewidth]{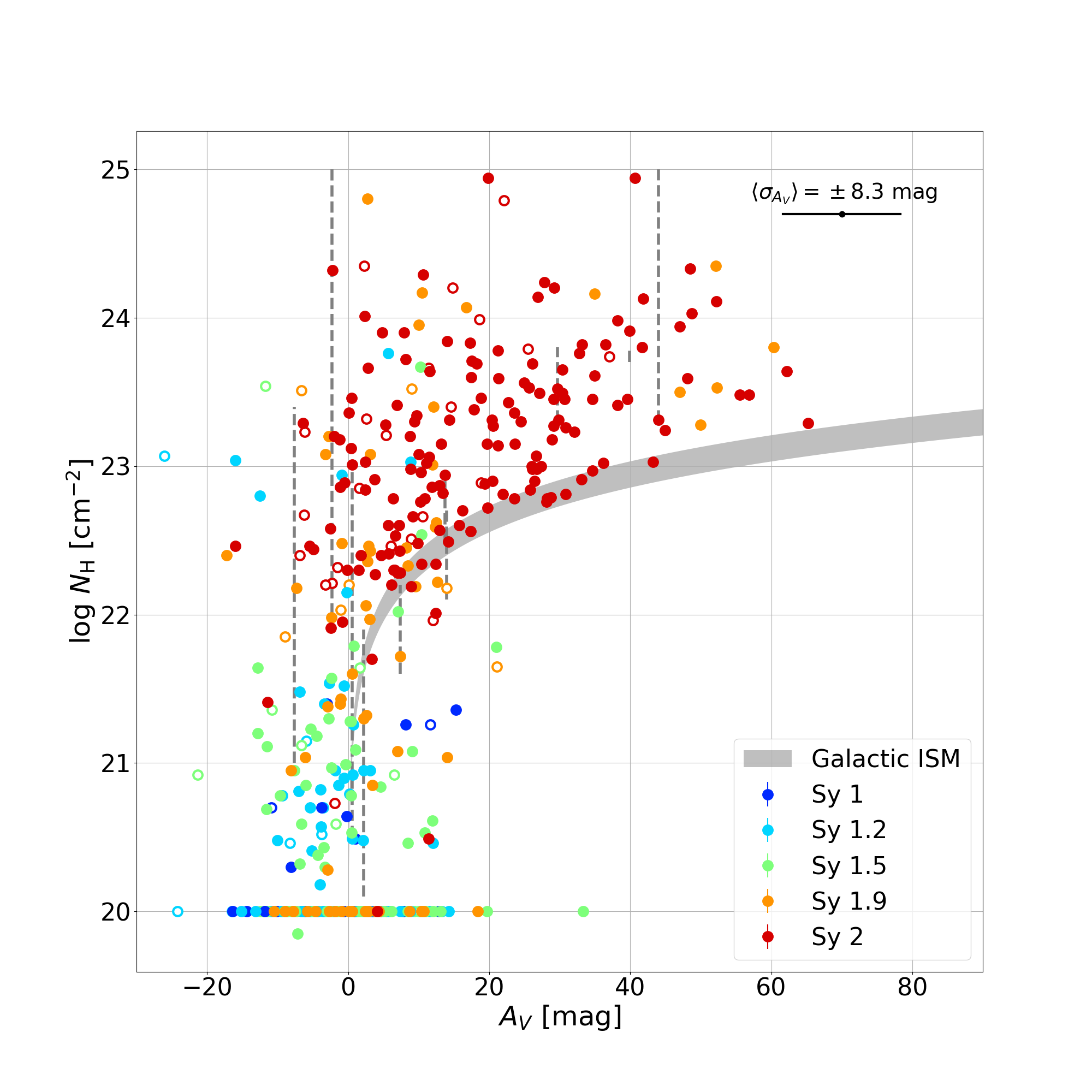}\par 
\caption{The comparison between the $A_V$ and $N_{\rm{H}}$ of our samples. We adopt the same colours as in Fig. \ref{fig:hist_final} for the same Seyfert types. The filled circles and open circles also represent the same samples as in Fig. \ref{fig:NH_beta}.  We show the typical $N_{\rm{H}}/A_V$ of the Galactic diffuse ISM with the grey band. The black segment in the upper right indicates the average of the $1\sigma$ error of the $A_V$ estimate, $\langle\sigma_{ A_V}\rangle=8.3$ mag. Dashed grey segments indicate the variable $N_{\rm{}H}$ for the common samples to Burtscher et al. (2016). The data for the variable $N_{\rm{}H}$ is obtained from Burtscher et al. (2016) and the citation therein.}
\label{fig:NH_A_V}
\end{figure*}

\begin{figure}
    \includegraphics[width=\columnwidth]{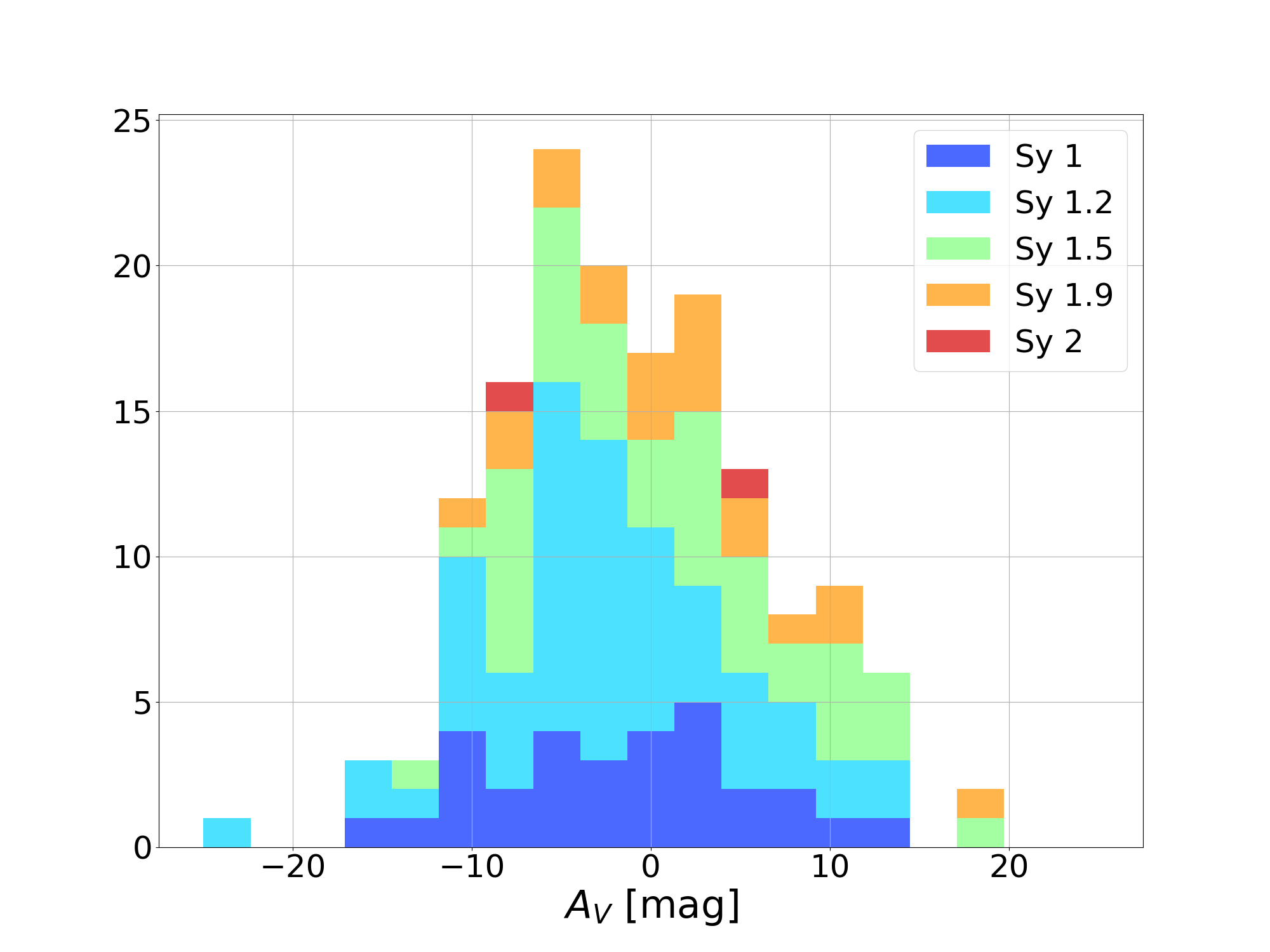}\par 
\caption{The histogram of the dust extinction $A_V$ for the samples with $\log N_{\mathrm{H}}\ [\mathrm{cm^{-2}}] \leq 20$. The colours indicate the Seyfert types of the samples in the same way as in Fig. \ref{fig:hist_final}. }
\label{fig:A_V_about0}
\end{figure}

Fig. \ref{fig:NH_A_V} shows the dust extinction compared with the hydrogen column density from the X-ray absorption.
As reported from Fig. \ref{fig:NH_beta}, almost all samples with $\log N_{\rm{H}}\ [\rm{cm^{-2}}]\lesssim22$ distribute around $A_V\sim0$ mag, and these targets are dominated by Seyfert type-1--1.5 AGNs.
Fig. \ref{fig:A_V_about0} shows the distribution of $A_V$ of samples with $\log N_{\mathrm{H}}\ [\mathrm{cm^{-2}}] \leq 20$.
The average and the standard deviation of them are calculated as $-0.74\pm 0.65$ mag and 8.0 mag, which would be consistent with a distribution of the average $A_V=0$ with the typical measurement error for them of $\langle\sigma_{A_V}\rangle(\log N_{\mathrm{H}}\leq20)=8.0$ mag.
Therefore, negative $A_V$ values for the same targets in Figs. \ref{fig:NH_A_V} and \ref{fig:A_V_about0} are considered to be mostly caused by the measurement error.

\color{black}
In contrast, the typical $A_V$ of the samples shows a broad correlation with $N_{\rm{H}}$ within $A_V\lesssim65$ mag in the range of $\log N_{\rm{H}}\ [\rm{cm^{-2}}]\gtrsim22$.
These samples with high $N_{\rm{}H}$ are dominated by type-1.9--2 AGNs. 
They show a target-to-target difference in $N_{\rm{H}}/A_V$ that spreads over about two orders of magnitude.
In Fig. \ref{fig:NH_A_V}, we draw the relation between $N_{\rm{}H}$ and $A_V$ for the standard $N_{\rm{}H}/A_V$ of the Galactic diffuse ISM \citep[$N_{\rm{H}}/A_V=(1.79$--$2.69)\times10^{21}\ \rm{cm}^{-2}\ mag^{-1}$,][]{Predehl95,Nowak12} for comparison.
The data points for obscured AGNs are located above the Galactic curve, which indicates that these obscured AGNs have typically larger $N_{\rm{H}}/A_V$ than the Galactic diffuse ISM. 
These trends for obscured AGNs agree with the results reported in the literature \citep[e.g.,][]{Maiolino01,Burtscher16}.
Although there are some data points below the Galactic curve at low $N_{\mathrm{H}}$ ($\log N_{\mathrm{H}}\lesssim22$), they are considered to be mostly consistent with small $A_V$ values on the Galactic curve with their measurement errors (see Fig. \ref{fig:A_V_about0}).
\color{black}
Furthermore, the lower envelope of the data distribution of obscured AGNs is almost consistent with the Galactic curve, which is reported by \cite{Burtscher16} for a smaller number of data.

\color{black}
In Fig. \ref{fig:NH_beta}, we draw the $N_{\rm{}H}-\beta$ relation for the standard Galactic diffuse ISM.
We found that the obscured AGNs with $\beta\lesssim0.65$ and $\log N_{\rm{}H}\ [\rm{}cm^{-2}]\gtrsim22$ have larger $N_{\rm{}H}$ than the expected from the NIR FVG assuming the Galactic diffuse ISM.
This trend is similar to that in Fig. \ref{fig:NH_A_V}.

\defcitealias{smith2014}{Paper~I}
\begin{table*}
 \caption{Main properties of our final samples and the results of our analyses.}
 \label{tab:result}
 \begin{tabular}{lllllllllll}
  \hline
  (1)&(2)&(3)$^a$&(4)$^a$&(5)$^a$&(6)&(7)&(8)&(9)$^b$&(10)&(11)$^b$\\
  Source & Counterpart & Redshift & Sy type & $\log N_{\rm{H}}$ & $r$ & $r_{\rm{NEO}}$ & $\beta$ & $\sigma_{\beta}$ & $A_V$ & $\sigma_{A_V}$ \\
  &&&&(cm$^{-2}$)&&&&&(mag)&(mag)\\
  \hline
SWIFTJ0001.0-0708 & 2MASX J00004876-0709117 & 0.0375 & 1.9 & 22.19 & 0.96 & 0.879 & 0.71 & 0.07 & 9.6 & 8.2 \\
SWIFTJ0001.6-7701 & 2MASX J00014596-7657144 & 0.0584 & 1.9 & 20.00 & --- & 0.866 & 1.04 & 0.09 & $-10.5$ & 8.0 \\
SWIFTJ0003.3+2737 & 2MASX J00032742+2739173 & 0.0397 & 2 & 22.86 & --- & 0.996 & 0.87 & 0.03 & $-1.1$ & 6.7 \\
SWIFTJ0006.2+2012 & Mrk 335 & 0.0258 & 1.2 & 20.48 & 0.981 & 0.943 & 1.03 & 0.06 & $-10.0$ & 7.5 \\
SWIFTJ0009.4-0037 & SDSS J000911.57-003654.7 & 0.0733 & 2 & 23.56 & --- & 0.922 & 0.52 & 0.08 & 25.0 & 9.8 \\
SWIFTJ0021.2-1909 & LEDA 1348 & 0.0956 & 1.9  &	21.98 & 0.98 & 0.977 & 0.89 & 0.05 & $-2.3$ & 6.8 \\
SWIFTJ0026.5-5308 & LEDA 433346 & 0.0629 & 1.9 & 20.00 & 0.983 & 0.982 & 0.93 & 0.06 & $-4.6$ & 7.1 \\
SWIFTJ0029.2+1319 & PG 0026+129 &	0.142 & 1.2 & 20.00 & 0.962 & 0.905 & 0.94 & 0.11 & $-4.6$ & 8.0 \\
SWIFTJ0034.6-0422 & 2MASX J00343284-0424117 & 0.213 & 2 & 23.45 & 0.992 & 0.892 & 0.41 & 0.06 & 30.7 & 8.2 \\
SWIFTJ0042.9-2332 & NGC 235A &  0.0222 & 1.9 & 23.5 & 0.939 & 0.942 & 0.36 & 0.04& 47.0 & 8.6 \\

  \hline
\multicolumn{10}{l}{\footnotesize $^a$ Data are taken from BASS AGN catalogue \citep{Koss17,Ricci17b}.}\\
\multicolumn{10}{l}{\footnotesize $^b$ The $1\sigma$ error.}\\
\multicolumn{10}{l}{\footnotesize (This table is available in its entirety in machine-readable form.)}\\
\end{tabular}
\end{table*}

\section{discussion}
\label{sec:discussion}

\subsection{Comparison of extinction estimates in different methods}
\label{subsec:comparison with previous studies}

\cite{Burtscher16} and \cite{Shimizu18} measured the dust extinction in different methods for some of our target AGNs.
In this section, we compare our estimates of dust extinction with those of these two studies.

\begin{figure}
 \includegraphics[width=\columnwidth]{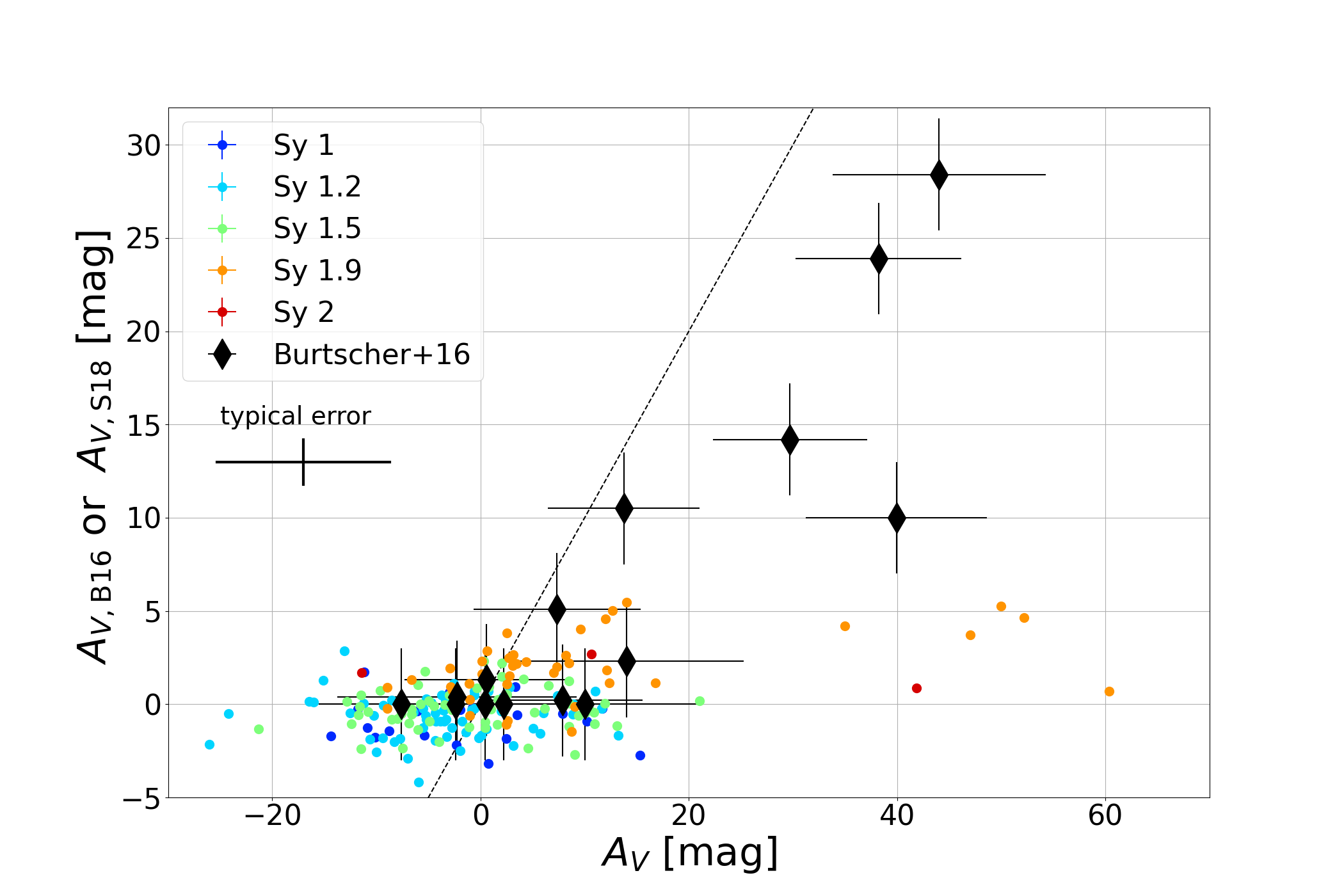}
 \caption{The comparison of our dust extinction estimates $A_V$ with those by Burtscher et al. (2016) ($A_{V,\rm{}B16}$) and Shimizu et al. (2018) ($A_{V,\rm{}S18}$).  The black diamonds with error bars represent the data for $A_{V,\rm{}B16}$ and $A_V$, the coloured dots represent the data for $A_{V,\rm{}S18}$ and $A_V$, and the black dashed line represents $A_V$ equals $A_{V,\rm{}B16}$ or $A_{V,\rm{}S18}$. The colours of dots indicate the different Seyfert types of the targets in the same way as in Fig. \ref{fig:hist_final}, and the error bars in the middle left represent the $\pm1\sigma$ error for the coloured dots.}
 \label{fig:A_V compare}
\end{figure}

\cite{Burtscher16} estimated the dust extinction of 29 AGNs based on the colour temperature of the $K$-band continuum emission from the unresolved AGN core. 
They obtained the amount of reddening of the NIR thermal emission from hot dust in the innermost dusty torus,  which is attributed to the dust extinction of the outer dusty torus.
In Fig. \ref{fig:A_V compare}, we compared our estimates of the dust extinction  with those of \cite{Burtscher16}, $A_{V,\rm{}B16}$ for the common 15 AGNs.
In Fig. \ref{fig:A_V compare}, $A_{V,\mathrm{B16}}$ gets larger as $A_V$ increases, and $A_V$ is consistent with $A_{V,\mathrm{B16}}$ for the 11 targets in $A_V<20$ mag.
However, for the four targets in $A_V>20$ mag, $A_V$ is systematically larger than $A_{V,\mathrm{B16}}$ by about 15 mag.


One possible reason for smaller $A_{V,\mathrm{B16}}$ than $A_V$ is the underestimation of the intrinsic colour temperature in the $K$ band assumed in \cite{Burtscher16}.
\cite{Burtscher16} adopted the intrinsic $K$-band colour temperature of $T=1311\pm129$ K, which is the mean colour temperature of their 13 type-1--1.9 AGNs.
This is somewhat lower than the intrinsic colour temperature suggested by $K-L$ or $H-K$ colour in other works \citep[e.g.,][]{Glass04,Kishimoto07}, although the $H-K$ colour may suggest a hotter colour temperature than that in the $K$ band.

\color{black} In the case of heavily obscured AGNs, \color{black} another possible reason is the effect of \color{black}the emission from cool extended dusty structure. \color{black} 
\cite{Burtscher16} assumed that the observed $K$-band emission came from the hot dust region in the innermost region of the dusty torus through the outer torus.
\color{black}However, recent IR interferometric observations of NGC 1068 \citep{Gamez_Rosas22} indicates that the $LM$-band emission come from cooler and extended structure shifted above the hot dust in the innermost dusty torus, and the $K$-band emission is also the same.
We consider that this interpretation would explain the difference between $A_V$ of \cite{Burtscher16} and those of this study: single-epoch NIR emission for heavily obscured AGNs may be contaminated by the cooler extended component, which causes relatively low dust extinction, while most of the time-variable flux component would come from the compact hot dust region in the innermost dusty torus, which causes relatively high dust extinction.
Furthermore, \color{black}although we use the NIR emission in the same way as \cite{Burtscher16}, the observed wavelengths of $W1$ and $W2$ bands are longer than that of the $K$ band, and so the former is more sensitive to the emission from the heavily obscured regions than the latter.
Consequently, the extinction estimates in this study are expected to be less affected by the \color{black} emission from such an extended structure than those in \cite{Burtscher16}, and the difference of $A_V$ occurs for heavily obscured AGNs. 

\color{black}
\cite{Shimizu18} estimated the dust extinction of the BASS AGNs by measuring the attenuation of the broad H$\alpha$ emission compared to the 14--150 keV hard X-ray emission, assuming that the luminosity ratio between them for unobscured AGNs is constant.
Because optical H$\alpha$ emission is more sensitive to dust extinction, their method is expected to be advantageous for measuring a small amount of dust extinction for less obscured AGNs, as they reported that the typical uncertainty was estimated as $\sigma_{A_V}=1.2$ mag.
In Fig. \ref{fig:A_V compare}, we compared our estimates of dust extinction $A_V$ with those of \cite{Shimizu18}, $A_{V,\rm{}S18}$.
For comparison, we calculated the $A_{V,\rm{}S18}$ for all of our targets whose broad H$\alpha$ luminosity is available.
We note that the value of $A_{V,\rm{}S18}$ presented in \cite{Shimizu18} is reproduced for common AGNs. 
We found that both estimates were roughly consistent with one another for $A_V\lesssim20\ \rm{}mag$ within the error.
However, $A_{V,\rm{}S18}$ was smaller than our estimates for $A_V\gtrsim20\ \rm{}mag$.
A similar trend was observed by \cite{Xu20} compared with their dust extinction estimates based on the mid-IR silicate absorption feature.
We suspect that $A_{V,\rm{}S18}$ for obscured AGNs is underestimated because of the contribution of the broad H$\alpha$ flux scattered in the polar region, as suggested by \cite{Xu20}.
\color{black}Although typical degrees of H$\alpha$ polarization for AGNs are very small \citep[$\sim1\%$, see e.g.][]{Ramos_Almeida16}, the scattered H$\alpha$ emission could be a significant contributor when the nuclear emission is attenuated by a factor of more than 100 ($\gtrsim5$ mag). 
\color{black}
This polarized H$\alpha$ emission is often observed for Seyfert type-2 AGNs by spectropolarimetry \citep[e.g.,][]{Antonucci85}.

\color{black}Although the uncertainty of our $A_V$ estimate is larger than that of \cite{Burtscher16} and \cite{Shimizu18}, our method can estimate $A_V$ for more-heavily obscured AGNs by using longer NIR wavelength, and can be applied to large number of AGNs easily by using the all-sky NIR monitoring data by \it{}WISE\rm{}.
These features of our method enables us to proceed statistical study of dust extinction for obscured AGNs.\color{black}

\subsection{Distribution of dust extinction to neutral hydrogen absorption}
\label{subsec:NH gap}

We here discuss possible origins for the excess of $N_{\mathrm{H}}/A_V$ ratios for obscured AGNs over the standard value for the Galactic diffuse ISM.
There are two possible explanations for the $N_{\mathrm{H}}/A_V$ excess over the Galactic value:
(1) the decrease of the $A_V$ caused by the depletion of the small dust grains \citep[e.g.][]{Maiolino01,Maiolino01b}, or (2) the increase of the $N_{\mathrm{H}}$ caused by dust-free gas clouds covering the line of sight in the broad-line region (BLR) of the AGN \citep[e.g.][]{Granato97,Burtscher16}.

In Scenario (1), the size distribution of dust grains in the dusty torus is larger than that of Galactic diffuse ISM, which decreases the observed dust reddening and extinction for fixed $N_{\rm{}H}$.
The lack of the 2175\AA\ carbon bump in the extinction curve of a typical type-1 AGN suggests the absence of small dust grains in AGN circumstances \citep[e.g.,][]{Maiolino01,Maiolino01b,Gaskell04,Czerny04}.
Two mechanisms are proposed for depleting small dust grains in the dusty torus: (i) the coagulation or aggregation of small dust grains and their growth into larger dust grains \citep[e.g.,][]{Maiolino01,Maiolino01b}, and (ii) the selective destruction of small dust grains because of dust sublimation \citep[e.g.,][]{Baskin18} or the Coulomb explosion \citep{Tazaki20}.

In Scenario (2), dust-free gas clouds are assumed to be located in the dust-free zone within the dust sublimation radius, such as in BLR, 
and their line-of-sight crossing causes a temporary increase in $N_{\rm{H}}$ without an increase in dust extinction.
This is supported by the time variation of the $N_{\rm{H}}$ in  time scales of months to years, which is observed in some AGNs \citep[e.g.,][]{Risaliti02,Burtscher16}.
\cite{Burtscher16} demonstrated the $N_{\rm{}H}$ time variation of 15 AGNs, which vary  between $22\lesssim\log N_{\rm{}H}\ [\rm{cm^{-2}}]\lesssim25$ in several years.
This fluctuating range roughly corresponds to the scattering range of obscured AGNs in Fig. \ref{fig:NH_A_V}, which strengthens this scenario.
Furthermore, the target-to-target difference of $N_{\rm{}H}/A_V$ for AGNs with $\log N_{\rm{}H}\ [\rm{}cm^{-2}]\gtrsim22$ can be explained by the difference in the extinction level because of the different gas density or number of dust-free gas clouds in the line of sight.

We here support the $N_{\rm{}H}$ excess of the dust-free gas clouds as a major origin to explain the behaviour of $N_{\rm{}H}/A_V$ of obscured AGNs, because (a) the dust-free gas scenario is easier to explain the large target-to-target difference in $N_{\rm{}H}/A_V$, 
(b) the extinction by the dusty torus with the Galactic dust properties can explain the lower envelope of the distribution of $N_{\mathrm{H}}/A_V$ data that is consistent with that of the Galactic diffuse ISM,
and (c) the dust-free gas scenario easily explains the $N_{\rm{}H}$ time variation.
For the common ten target AGNs with \cite{Burtscher16}, we overwrite the range of time variation of $N_{\rm{}H}$ in Fig. \ref{fig:NH_A_V} \citep[][ and citation therein]{Burtscher16}.
The amplitude of $N_{\rm{}H}$ variation is comparable to the $N_{\rm{}H}$ range of the distribution of the obscured AGNs.

In addition, large dust grains may influence the $N_{\mathrm{H}}/A_V$ distribution in a minor manner.
We note that the lower envelope of the distribution of obscured AGNs in the $A_V\gtrsim40$ mag region may appear to move slightly upward from that in the $A_V\lesssim40$ mag region in Fig. \ref{fig:NH_A_V}.
This might indicate the effect of large dust grains in the mid-plane region of the dusty torus that might be observed from an edge-on view.
\cite{Wada19,Wada21} suggest that the emission from the central engine does not effectively reach the mid-plane of the dusty torus and that the snow line is located around several parsecs from the centre.
They concluded that, in such a case, small ice grains might grow into larger dust grains through aggregation.

\section{Conclusion}
\label{sec:conclusion}

In this study, we estimated the dust extinction of X-ray-selected AGNs using long-term NIR monitoring data obtained by \it{}WISE\rm{}, and examined the relationship between the dust extinction and hydrogen column density.
Our results and conclusions are as follows:\\

1. We measured the flux variation gradient (FVG) using the flux data in $W1$ and $W2$ bands ($\equiv\beta$) of 513 samples and obtained it with an uncertainty of $\sigma_{\beta}<0.2$ for more than 90\% (463/513) of these samples.\\

2. We compared the NIR FVGs with Seyfert types and the line-of-sight neutral hydrogen column density $N_{\rm{}H}$.
For AGNs with $\log N_{\rm{H}}\ [\rm{cm^{-2}}]\lesssim22$, the majority of which are Seyfert type 1--1.5 AGNs, the NIR FVGs are distributed in a relatively narrow range regardless of the Seyfert types and the $N_{\rm{}H}$.
However, for AGNs with $\log N_{\rm{H}}\ [\rm{cm^{-2}}]\gtrsim22$, the majority of which are Seyfert type 1.9--2 AGNs, the NIR FVGs show clear reddening when $N_{\rm{}H}$ increases.\\

3. We determined the intrinsic NIR FVG by fitting the NIR FVGs for unobscured AGNs, and then we calculated the dust extinction for the 463 AGNs based on the reddening of their NIR FVGs.
This novel method is advantageous for measuring the dust extinction for a large sample of obscured AGNs because it uses the variable flux component attributed to the AGN, allows us to ignore the correction for host-galaxy emission, and because the thermal emission of hot dust in the NIR is observable even for obscured AGNs.\\

4. The $N_{\rm{H}}/A_V$ of our obscured AGN samples is typically greater than that of the Galactic diffuse ISM, and the lower envelope of the distribution of these $N_{\rm{H}}/A_V$ is consistent with that of the Galactic diffuse ISM.
They have target-to-target scatter that spans approximately two orders of magnitude as has been reported in the literature.
These behaviours of obscured AGNs can be explained by the scenario in which dust-free gas clouds in the BLR crossing the line of sight cause both the increase of $N_{\rm{}N}$ and its time variation.

\section*{Acknowledgements}
SM is supported by JST SPRING, Grant Number JPMJSP2108.
HN is supported by Japan Society for the Promotion of Science (JSPS) KAKENHI with the Grant number of 19K21884, 20H01941, and 20H01947.
HS is supported by Japan Society for the Promotion of Science (JSPS) KAKENHI with the Grant number of 19K03917.

This publication has made use of data products from the \it{}Wide- field Infrared Survey Explorer\rm{}, which is a joint project of the University of California, Los Angeles, and the Jet Propulsion Laboratory/California Institute of Technology, funded by the National Aeronautics and Space Administration. This publication also makes use of data products from \it{}NEOWISE\rm{}, which is a project of the Jet Propulsion Laboratory/California Institute of Technology, funded by the Planetary Science Division of the National Aeronautics and Space Administration.

We also acknowledge the use of public data from the BAT AGN Spectroscopic Survey.

This research has made use of the NASA/IPAC Extragalactic Database (NED), which is operated by the Jet Propulsion Laboratory, California Institute of Technology, under contract with the National Aeronautics and Space Administration.

\section*{data availability}
The \it{}WISE\rm{} data used in this study are publicly available in the NASA/IPAC Infrared Science Archive (\url{https://irsa.ipac.caltech.edu/Missions/wise.html}). 
The data of BASS AGN catalogue used in this study are also publicly available from the BASS website (\url{https://www.bass-survey.com}).
The all data derived by the analyses of flux-flux plots in this study are available in the online version of this paper.   



\bibliographystyle{mnras}
\bibliography{WISE_1} 







\bsp	
\label{lastpage}
\end{document}